\documentstyle[twocolumn,seceq]{jpsj}

\def\runtitle
{
Spin Polarization and Magneto-Coulomb Oscillations 
}
\def\runauthor {K. {\sc Ono}, H. {\sc Shimada} and Y. {\sc Ootuka}}

\title
{
Spin Polarization and Magneto-Coulomb Oscillations \\
in Ferromagnetic Single Electron Devices
}

\author
{ 
Keiji {\sc Ono}\footnote{E-mail: ono@crc.u-tokyo.ac.jp}, Hiroshi 
{\sc Shimada} and Youiti {\sc Ootuka}
}

\inst
{
Cryogenic Center, University of Tokyo, Yayoi 2-11-16, Bunkyo-ku, Tokyo 113-0032
}

\recdate
{
\today
}

\abst
{
     The magneto-Coulomb oscillation, the single electron repopulation induced by external magnetic field, observed in a ferromagnetic single electron transistor is further examined in various ferromagnetic single electron devices. In case of double- and triple-junction devices made of Ni and Co electrodes, the single electron repopulation always occurs from Ni to Co electrodes with increasing a magnetic field, irrespective of the configurations of the electrodes. The period of the magneto-Coulomb oscillation is proportional to the single electron charging energy. All these features are consistently explained by the mechanism that the Zeeman effect induces changes of the Fermi energy of the ferromagnetic metal having a non-zero spin polarizations. Experimentally determined spin polarizations are negative for both Ni and Co and the magnitude is larger for Ni than Co as expected from band calculations. 
}

\kword
{
single electron charging effect, Coulomb oscillation, ferromagnetic tunnel junction, electron spin-polarization
}

\begin{document}
\sloppy
\maketitle

\section{Introduction}
     There have been intensive studies on the single electron charging effects last several years.~\cite{AvrLik, SCT} It is known that the single electron charging effect couples to other degrees of freedom introduced by using exotic materials as electrodes such as superconductors or semiconductors, and gives interesting phenomena. The single electron devices made of ferromagnetic electrodes are in growing interest in which the spin-polarized electrons could add more fertile aspects to the physics of the single electron devices.~\cite{Ono2D, OnoSET, Co-dots, OsciMR, Takahashi, Sim}

     Recently, we found two novel phenomena in a Ni/Co/Ni SET, a single electron transistor made of Co island and Ni leads electrodes. One is the magneto-Coulomb oscillation (MCO), i.e., single electron repopulation by a magnetic field even in a fixed gate voltage, and the other is the enhancement of the tunnel magneto-resistance in the Coulomb blockade regime.~\cite{OnoSET} Concerning the former, we have proposed a model that is based on a magnetic-field-induced change in the Fermi energy (chemical potential) of the spin-polarized electron systems in ferromagnetic electrodes.~\cite{Sim} According to the model, SETs with different configuration of electrodes should show MCOs in a different way. In this paper, we present experimental results of various systems such as a Co/Ni/Co SET, Al/Co/Al SETs and a Ni/Co/Ni/Co triple junction, as well as the previously reported result of the Ni/Co/Ni SET. We found the model consistently explains the observed MCOs in these systems. In $\S$2, we describe the mechanism of MCO in ferromagnetic SET. In $\S$3, we present the observed MCOs in the various systems and compare with the model in $\S$2. The conclusion is described in $\S$4.

\section{ Magneto-Coulomb oscillation } 
     In this section, we briefly explain the model for MCO.~\cite{Sim} Throughout this paper, we deal a high magnetic field where all magnetizations of the electrodes are parallel to the magnetic field.

     In transition metal ferromagnets, there is a significant difference in the density of states for majority ($D\rm{_{+}}$) and minority ($D\rm{_{-}}$) spins at the Fermi level due to the exchange splitting in d-bands. According to the band calculations, the net spin-polarizations defined as
\begin{equation}
\label{MCO1}
P=\frac{ D{\rm_{+}}-D{\rm_{-}} }{ D{\rm_{+}}+D{\rm_{-}} }
\end{equation}
are around $-81 \sim -78 {\rm \%}$ for Ni and $-64  \sim -43 {\rm \%}$ for Co.~\cite{Band1, Band2, Band3, Band4, Band5} They are in remarkable contrast with the spin polarizations for $tunneling$ electrons.~\cite{SPT} The latter differ from $P$'s even in the signs because the tunneling electrons are dominated by that from the s-d hybridized bands.~\cite{Stearns}

     Suppose that an isolated ferromagnet of spin polarization $P$ is in a magnetic field $H$. Then the energy bands for the majority (minority) spin is shifted by an amount of $- (1/2)g\mu{\rm_{B}}H$ ($(1/2)g\mu{\rm_{B}}H$) due to the Zeeman effect, where $g$ and $\mu\rm{_{B}}$ are the g-factor and the Bohr magneton, respectively. After repopulation of the electrons between the majority and minority bands, the Fermi energy (chemical potential) $\varepsilon{\rm_{F}}$ will be shifted by an amount of
\begin{equation}
\label{MCO2}
\Delta\varepsilon {\rm_{F}} = -\frac{1}{2}P g \mu{\rm_{B}} H 
\end{equation}
due to the difference in the majority and minority density of states at $\varepsilon\rm{_{F}}$.

     If we consider a SET with a ferromagnetic island and nonmagnetic ($P$ = 0) leads, the island is still regarded as $isolated$ as long as $|\Delta\varepsilon{\rm_{F}}| < E\rm{_{C}}$ because the number of electrons on the island is almost fixed due to Coulomb blockade, where $E\rm{_{C}}$ is the single electron charging energy of the island. When $|\Delta\varepsilon{\rm_{F}}|$ exceeds $E\rm{_{C}}$, the mean electron number of the island changes by one. In this way, the magnetic field switches the SET periodically and leads to MCO with the oscillation period of the field $\Delta H$ given as
\begin{equation}
\label{MCO3}
\frac{1}{2} |P| g \mu{\rm_{B}} \Delta H = 2E{\rm_{C}}.
\end{equation}
The mean electron number of the island increases (decreases) with increasing $|H|$ for positive (negative)$P$.

     Next, we consider a SET with nonmagnetic island and ferromagnetic leads, where the leads are connected to grounded nonmagnetic contact pads. By applying the magnetic field, repopulation of charges occurs at the interfaces between the leads and the pads in order to match their electrochemical potentials $\varepsilon{\rm_{F}} - e\phi$ where $\phi$ is the electrostatic potential of the electrode. As a result, the electrostatic potential energy in the leads $-e \phi{\rm_{lead}}$ changes by an amount $(1/2)Pg\mu{\rm_{B}}H$ because the electrochemical potentials of the contact pads does not change. In case that the total self-capacitance of the island is dominated by the junction capacitances, as is the usual for metallic SETs, the electrostatic potential energy in the island $-e \phi{\rm_{island}}$ will also change by the same amount as those in the leads. This gives the same effect on the single electron transport as changing $-e \phi{\rm_{island}}$ by the gate voltage does, and leads to MCO with the oscillation period of the magnetic field given as eq.~(\ref{MCO3}). But in this case, the mean electron number of the island decreases (increases) with increasing $|H|$ for positive (negative) $P$.

     For a SET consists of ferromagnetic island ($P = P\rm{_{island}}$), ferromagnetic leads ($P = P\rm{_{lead}}$) and nonmagnetic contact pads ($P = 0$, $\phi = 0$) to which the leads are connected, the two effects described above should be combined. The period of MCO is then given as
\begin{equation}
\label{MCO4}
\frac{1}{2} |P{\rm_{island}}-P{\rm_{lead}}| g \mu{\rm_{B}} \Delta H  = 2E{\rm_{C}}.
\end{equation}
The mean electron number of the island increases with increasing $|H|$ for the case of $P\rm{_{lead}} < P\rm{_{island}}$ and vice versa. 

\section{ Experiment } 
     The Ni/Co/Ni SET, the Co/Ni/Co SET and the Ni/Co/Ni/Co triple junction are consist of small (junction area $\approx$ 0.01 $\mu\rm{m}^2$) Ni/NiO/Co tunnel junctions in series that are fabricated on Si substrates using a deposition mask made of Si$_3$N$_4$ membrane.~\cite{OnoFab} Nickel and cobalt electrodes are thermally evaporated at a pressure lower than 2 $\mu\rm{Torr}$. Oxidation of Ni is performed by use of oxygen plasma of 80 mTorr for 3 min. The Al/Co/Al SETs consist of Al/Al$_2$O$_3$/Co junctions. Al$_2$O$_3$ tunnel barriers are formed in oxygen of 80 mTorr for 3 min without the plasma. All lead electrodes of the samples are connected to contact pads made of Au thin films. Gate electrodes made of silver paste are located on the backside of the substrates. The geometry of the working part of the samples are shown in Figs.~\ref{NiCo}, \ref{AlCo} and \ref{Trip}. All samples except the Co/Ni/Co SET are covered with SiO 50 nm thick to prevent further oxidation of the electrodes. Detailed characterizations of the samples are described in following subsections. The samples are cooled down to around 20 mK in a dilution refrigerator which is equipped with a 8 T superconducting magnet. Magnetic field is applied parallel to the long axis of the electrodes.

\subsection{ Ni/Co/Ni SET } 
     The Ni/Co/Ni SET consists of Co island (about 150-nm wide, 14-nm thick and 2.5 $\mu\rm{m}$ long), and Ni leads (about 250-nm wide, 13-nm thick). The capacitance between the island and the gate $C\rm{_{G}}$ is determined from the period of the Coulomb oscillation (CO) and is about 0.6 aF, which is much smaller than the expected junction capacitances (order of 1 fF). The tunnel resistance of each junction $R\rm{_{T}}$ is determined from the half of the sample resistance at 4.2 K to be 35 k$\Omega$. The charging energy $E\rm{_{C}}$ of the island is estimated from the maximum threshold voltage. But, because of the smeared threshold characteristics, only a crude estimation is allowed for this sample, that is, $E\rm{_{C}} = 25 \sim 50 \mu\rm{eV}$.

     Figure \ref{NiCo}~(a) shows gray-scale plot of the zero-bias resistance measured during raster scanning the magnetic field and the gate voltage $V\rm{_{G}}$. Bright region corresponds to high resistance. It is seen that almost continuous and monotonic shift of the phase of CO toward the negative $V\rm{_{G}}$ direction upon application of the magnetic field, which is found to be independent of the direction and the sweep directions of the magnetic field except at low field ($|H| < 10 {\rm kOe}$). We also find the slope of the equi-phase lines is temperature independent at least below about 0.4 K where the CO is observable. The negative slope of the equi-phase lines means, for fixed gate voltage, the mean excess electron number in the island increases one by one with increasing magnetic field (MCO). According to the model in $\S$2, it is the case where $P\rm{_{lead}} < P\rm{_{island}}$, and is consistent with the band calculations where $P{\rm_{Ni}} < P{\rm_{Co}} < 0$. Quantitatively, eq.~(\ref{MCO4}) gives $P{\rm_{Co}}-P{\rm_{Ni}}= +29 \sim +86 \rm{\%}$ when we use the experimentally determined values; $2E{\rm_{C}} = 50 \sim 100 \mu\rm{eV}$, $\Delta H = 20 \sim 30 \rm{kOe}$ and assume $g = 2$. This value is not far from the theoretical value, $P{\rm_{Co}}-P{\rm_{Ni}}= +14 \sim +38 \rm{\%}$.

\subsection{ Co/Ni/Co SET } 
     When we exchange the materials used for the island and the leads, i.e., in SETs with Ni island and Co leads, the mean electron number will $decrease$ with increasing $|H|$, i.e., the CO-phase should shift toward the positive $V\rm{_{G}}$ direction by applying the magnetic field. In order to confirm this, we fabricated the Co/Ni/Co SET. The Ni island electrode is about 150 nm wide, 13 nm thick and 2.5 $\mu\rm{m}$ long and the Co lead electrodes are about 250 nm wide, 14 nm thick. The tunnel resistance of each junction is about 15 $\rm{k}\Omega$ determined from the half of the sample resistance at 4.2 K. The charging energy is difficult to determine due to the low resistance. But we think they do not differ much from that of the Ni/Co/Ni SET because they have similar junction areas.

     Figure \ref{NiCo}~(b) exhibits gray-scale plot of the zero-bias conductance of the Co/Ni/Co SET. The CO-phase clearly shifts to the positive $V\rm{_{G}}$ direction by applying the field, which is consistent with the expectation. The equi-phase lines are, however, not so smooth as that of the Ni/Co/Ni SET and details of them do not reproduce in different runs of sweeping the field. Such fluctuation of the CO-phase is frequently ascribed to the fluctuation of the background charges. In the present systems, however, we can mention another cause; the absence of SiO passivation layer. A thick surface oxide layer formed in air can pin the surface spins, induce the inhomogeneity of magnetization, and affect the magnetic-field-induced shift of the Fermi energy.

\subsection{ Al/Co/Al SET } 
     We also investigate MCO in a ferromagnet/nonmagnet system. We chose Al for nonmagnetic material because of its easy-to-form stable oxides Al$_2$O$_3$ which makes it possible to fabricate high-$R\rm{_{T}}$ SET in which $E\rm{_{C}}$ can be determined accurately from the threshold characteristics. Drawback of using Al is that the measurements are restricted only at high fields where the superconductivity of Al is quenched.  We fabricated two Al/Co/Al SET samples labeled \#1 and \#2 below. Al/Co/Al SETs consist of Co island (about 150-nm wide, 15-nm thick) and Al leads (about 250-nm wide, 15-nm thick). The lengths of the electrodes are 2.5 $\mu\rm{m}$ for \#1 and 2.0 $\mu\rm{m}$ for \#2. $R\rm{_{T}}$'s determined by the differential resistance at high bias are 80 $\rm{k}\Omega$ (\#1) and 255 $\rm{k}\Omega$ (\#2). $E\rm{_{C}}$'s are determined from both the threshold voltage and the offset voltage characteristics and are about 50 $\mu\rm{eV}$ and 110 $\mu\rm{eV}$ for \#1 and \#2 respectively.

     Figure \ref{AlCo}~(a) and \ref{AlCo}~(b) show the gray-scale plots of the threshold voltages for the samples \#1 and \#2 respectively. The absence of data for $H < 20 \rm{kOe}$ is due to the superconductivity of Al where superconducting gap is added to the threshold voltage. The CO-phase shifts to the positive $V\rm{_{G}}$ directions for both samples, which is consistent with the negative $P\rm{_{Co}}$. The slope of the equi-phase line is larger ($\Delta H$ is smaller) for \#1 than for \#2. In Fig. \ref{AlCo}~(c), we plot $(\mu{\rm_{B}}\Delta H)^{-1}$ against $(2E\rm{_{C}})^{-1}$. Due to the stochastic small jumps of the CO-phase, $(\mu{\rm_{B}}\Delta H)^{-1}$'s are slightly different for three different runs of sweeping the magnetic field. We see $(\mu{\rm_{B}}\Delta H)^{-1}$ is proportional to $(2E\rm{_{C}})^{-1}$ as expected from eq.~(\ref{MCO3}). The ratio of the two quantity is $|P{\rm_{Co}}|$ if we assume $g = 2$. The doted line in the figure corresponds to $P\rm{_{Co}} \approx -37 \rm{\%}$ which is in fairly good agreement with the results of the band calculations $-64 \sim -43 \rm{\%}$. 

\subsection{ Ni/Co/Ni/Co Triple Junction } 
     Triple junction system of Ni/NiO/Co consists of Ni electrodes (about 250-nm wide, 13-nm thick) and Co electrodes (about 150-nm wide, 14-nm thick). Length of the islands are 2.5 $\mu\rm{m}$. $R\rm{_{T}}$ estimated from 1/3 of the sample resistance at 4.2 K is 55 $\rm{k}\Omega$.

     Figure \ref{Trip}~(a) shows the gate voltage dependence of the zero-bias conductance $G$ of the Ni/Co/Ni/Co triple junction. The observed oscillations have two distinct features: (1) There is a large-scale modulation of the oscillation with a period of the gate voltage of about 1.5 V. (2) Each peak of the oscillation is split into two, which is most evident near the antinodes of the modulation. The magnetic field dependence of the oscillation is plotted in Fig. \ref{Trip}~(b) for $H > 20 \rm{kOe}$ where the reproducible characteristic is obtained. The bright region corresponds to high conductance. The plot shows that the phase of the modulation shifts toward the positive $V\rm{_{G}}$ direction with increasing field roughly in a monotonic manner. On the other hand, the position of each split peak does not change in the field except when they pass through the nodes of the modulation. Note the fact that the magnetic field changes the shape of the $G$-$V\rm{_{G}}$ curve, which is in good contrast with the case of the SETs where the $G$-$V\rm{_{G}}$ curves are only shifted along the $V\rm{_{G}}$ axis by the magnetic field.

     In triple junction systems with two independent gates which couples to two islands respectively (the single electron pumps), the ground state configuration of mean excess electron numbers for the islands, ($N\rm{_{1}}$, $N\rm{_{2}}$), is determined by $n{_i} = C{\rm_{G}}{_i}V{\rm_{G}}{_i}/e$ ($i$ = 1, 2). Figure \ref{Trip}~(c) shows the ($N\rm{_{1}}$, $N\rm{_{2}}$) diagram in ($n\rm{_{1}}$, $n\rm{_{2}}$)-plane for zero-bias voltage. Here, we assume identical junction capacitances $C$ for three junctions and small gate capacitances ($C \gg C{\rm_{G}}{_i}$).~\cite{SCT} If one changes the gate voltages in the manner $n{\rm_{1}}=n{\rm_{2}}$, the CO with split peaks will be obtained. In our triple junctions, a single gate electrode couples to both Co island and Ni island through $C\rm{_{G1}}$ and $C\rm{_{G2}}$ respectively, and $C\rm{_{G2}}$ is expected to be somewhat larger than $C\rm{_{G1}}$ because the width of the Ni island is larger than that of Co island. Thus sweeping the gate voltage corresponds to traversing the $n{\rm_{1}}$-$n{\rm_{2}}$ diagram along the line with a slope somewhat larger than unity, $n{\rm_{2}} = (C{\rm_{G2}}/C{\rm_{G1}}) n{\rm_{1}}$, as shown in the figure, which results in the modulation of the peak-split CO. In order to analyze the moduration characteristics, we assume the conductance $G(n{\rm_{1}}, n{\rm_{2}})$ is given as
\begin{equation}
\label{TRIP1}
G(n{\rm_{1}}, n{\rm_{2}}) \propto -\cos 2\pi( n{\rm_{1}} - n{\rm_{01}} )
                      -\cos 2\pi( n{\rm_{2}} - n{\rm_{02}} ) + {\rm const.}
\end{equation}
where $n{\rm_{0}}{_i}$ ($i$ = 1, 2) is the offset charge on each island. The expression will be valid for weak Coulomb blockade regime with negligible inter-island capacitance. The observed $G$-$V\rm{_{G}}$ curve are fitted well by eq.~(\ref{TRIP1}) except for the peak-splitting which is a result of the capacitive coupling between the islands. $C\rm{_{G1}}$ and $C\rm{_{G2}}$ obtained by the fitting are 0.49 aF and 0.60 aF respectively, which are reasonable values compared to other SETs.

     Next, we examine the effect of the magnetic field. According to the experiments on the Ni/Co/Ni SET and the Co/Ni/Co SET, the mean electron number of Ni (Co) island decreases (increases) with increasing magnetic field in the Ni/Co/Ni/Co triple junction. This means that applying the magnetic field corresponds to traversing the $n_{1}$-$n_{2}$ diagram along the line $n_{2} = - n_{1}$ as shown in Fig. \ref{Trip}~(c). Thus, one can scan over the $n_{1}$-$n_{2}$ diagram using the gate voltage and the magnetic field, instead of using two independent gates. Figure \ref{Trip}~(d) shows gray-scale $V{\rm_{G}}$-$H$ diagram obtained using eq.~(\ref{TRIP1}) with the fitted $C{\rm{_G}}{_i}$'s. $\Delta H$ is the period of MCO where $\Delta N_{1}=-\Delta N_{2}=1$. The figure exhibits the shift of the modulation toward the positive $V\rm{_{G}}$ direction while the each peak position of the oscillation does not shift in the magnetic field until they pass through the nodes of the modulation, which reproduce the observed result. The observed period of the field $\Delta H$ is around 50 kOe, which is larger than that of the Ni/Co/Ni SET by a factor 1.6 $\sim$ 2.5 and is reasonable considering the smaller junction areas of the Ni/Co/Ni/Co triple junction.

\section{ Conclusion } 
     We have found MCOs in various ferromagnetic or partially ferromagnetic single electron devices such as the Ni/Co/Ni SET, the Co/Ni/Co SET, the Al/Co/Al SETs and the Ni/Co/Ni/Co triple junction. We observed that, with increasing the magnetic field, the electrons ware repopulated from Ni electrodes to Co electrodes and from Co electrodes to Al electrodes. We found that the period of MCO was proportional to the charging energy in the Al/Co/Al SETs. All these features are consistently explained by the mechanism that the Zeeman effect induces changes of the Fermi energy of the ferromagnetic metal having a non-zero spin polarizations $P$, and by the relation $P{\rm_{Ni}} < P{\rm_{Co}} < P{\rm_{Al}} = 0$ from the band calculations. From the experimental data, we obtain $P{\rm_{Co}}-P{\rm_{Ni}}= +29 \sim +86 \rm{\%}$ and $P{\rm_{Co}} \approx -37 \rm{\%}$. These values are in fairly good agreement with the values from the band theories, $P{\rm_{Co}}-P{\rm_{Ni}}= +14 \sim +38 \rm{\%}$ and $P{\rm_{Co}} \approx -64 \sim -43 \rm{\%}$.

     This work was supported by CREST Project of Japan Science and Technology Corporation, and by Grant-in Aid for Scientific Research Project on Priority Area from the Ministry of Education, Science, Sports and Culture of Japan.

\begin{figure}

\caption
{
Gray-scale plots of the zero-bias resistance or conductance for (a) the Ni/Co/Ni SET and (b) the Co/Ni/Co SET. The slopes of the equi-phase lines of the Coulomb oscillations are negative for the Ni/Co/Ni SET and positive for the Co/Ni/Co SET, which indicate that, by the application of the magnetic field, the electrons repopulate from the Ni electrodes to the Co electrodes in both SETs. 
}
\label{NiCo}
\end{figure}

\begin{figure}

\caption
{
(a), (b): Gray-scale plots of the threshold voltages of two Al/Co/Al SETs which have different $E\rm{_{C}}$'s. Phase of the Coulomb oscillation shifts to the positive $V\rm{_{G}}$ directions in both samples. The slope of the equi-phase lines is larger for smaller $E\rm{_{C}}$. (c) Liner relation between the period of the magneto-Coulomb oscillation $\Delta H$ and the charging energy $E\rm{_{C}}$, which is expected from eq.~(\ref{MCO3}). 
}
\label{AlCo}
\end{figure}

\begin{figure}

\caption
{
(a) Gate voltage dependence of zero-bias conductance of Ni/Co/Ni/Co triple junction which shows peak-split Coulomb oscillation with large scale modulation. (b) Gray-scale plot of the zero-bias conductance. The plot shows that the phase of the modulation shifts toward the positive $V\rm{_{G}}$ direction with increasing magnetic field while the position of each split peak does not change in the field except when they pass through the nodes of the modulation. (c) Ground state configuration of mean excess electron numbers for two islands ($N\rm{_{1}}$, $N\rm{_{2}}$) for given $n{_i} = C{\rm_{G}}{_i}V{\rm_{G}}{_i}/e$ ($i$ = 1, 2). Gate-voltage-induced and magnetic-field-induced changes on ($N\rm{_{1}}$, $N\rm{_{2}}$)'s are denoted by arrows. (d) Gray-scale $V{\rm_{G}}$-$H$ diagram using eq.~(\ref{TRIP1}). $\Delta H$ is the period of MCO where $\Delta N{\rm_{1}}=-\Delta N{\rm_{2}}=1$. 
}
\label{Trip}
\end{figure}

\end{document}